# Self-discharge mitigation in a liquid metal displacement battery

Kashif Mushtaq[a,b,c,*], Ji Zhao[a], Norbert Weber[a,d], Adelio Mendes[b], Donald R. Sadoway[a]

[a] Department of Materials Science and Engineering, Massachusetts Institute of Technology, 77 Massachusetts Avenue, Cambridge, MA 02139-4307, United States

[b] LEPABE - Laboratory for Process Engineering, Environment, Biotechnology and Energy, Faculty of Engineering, University of Porto, Rua Dr. Roberto Frias, 4200-465 Porto, Portugal

[c] Department of Mechanical Engineering, School of Mechanical and Manufacturing Engineering, National University of Sciences and Technology, H-12, Islamabad, Pakistan

[d] Helmholtz-Zentrum Dresden-Rossendorf, Bautzner Landstr. 400, 01328 Dresden, Germany

***Corresponding author**.

*E-mail address*: kashifmushtaq@outlook.com (K. Mushtaq)

**Abstract**

Recently, a disruptive idea was reported about the discovery of a new type of battery named Liquid Displacement Battery (LDB) comprising liquid metal electrodes and molten salt electrolyte. This cell featured a novel concept of a porous electronically conductive faradaic membrane instead of the traditional ion-selective ceramic membrane. LDBs are attractive for stationary storage applications but need mitigation against self-discharge. In the instant battery chemistry, Li|LiCl-$PbCl_2$|Pb, reducing the diffusion coefficient of lead ions can be a way forward and a solution can be the addition of PbO to the electrolyte. The latter acts as a supplementary barrier and complements the function of the faradaic membrane. The remedial actions improved the cell's coulombic efficiency from 92% to 97% without affecting the voltage efficiency. In addition, the limiting current density of a 500 mAh cell increased from 575 to 831 mA cm$^{-2}$ and the limiting power from 2.53 to 3.66 W. Finally, the effect of PbO on the impedance and polarization of the cell was also studied.

# 1. Introduction

Climate change is one of the biggest challenges of the current era. Europe is committed to achieving the vision of a climate-neutral society by 2050 [1]. This transition motivates the search for innovative solutions that result in the mitigation of the adverse effects of energy consumption and conversion. This also drives how electrical energy is being produced, consumed, and stored. Implementation of electric vehicles is also a key step toward battling $CO_2$ emissions. Batteries are the true enablers for this vision, but these need to be made safer, better performing, more sustainable, and affordable [2]. Batteries should exhibit high performance beyond their capabilities today. Furthermore, reliable batteries can provide greater resiliency to the existing grid while also lowering emissions [3]. Battery chemistries that have been deployed in stationary storage applications include lithium-ion, lead-acid, redox-flow, and high-temperature batteries such as sodium-sulfur and sodium-nickel chloride (ZEBRA) [4].

The transition to a more electricity-based society needs disruptive ideas that can enable the creation of sustainable batteries at the price point of the market. In line with these efforts Group Sadoway laboratory at MIT has reported the discovery of a new type of battery named liquid displacement battery (LDB) [3] which exploits the advantages of the sodium-nickel chloride electrochemistry without requiring the use of the β''-$Al_2O_3$ Na-ion conducting membrane. The fragility and brittleness of the β''-$Al_2O_3$ ceramic set severe design constraints and limit its performance. Also, its vulnerability to attack by transition-metal ions in a chloroaluminate melt is a concern. LDB has a new concept of a porous electronically conductive membrane instead of an ion-selective ceramic conductor. This membrane removes the barrier for researchers to choose only liquid Na for the negative electrode and solid transition-metal halides for the positive electrode [3].

The schematic of a liquid displacement battery based on the electrochemistry of Li|LiCl-$PbCl_2$|Pb is shown in Fig. 1. The negative electrode is comprised of liquid Li-Pb, where Li is the active species and Pb serves as host metal. The eutectic composition of LiCl-KCl is used as a molten-salt electrolyte. Finally, liquid Pb is serving as the positive electrode and $PbCl_2$ exists in the electrolyte in a dissolved state. The electrochemistry of Pb/$Pb^{2+}$ displacement reactions in molten LiCl-KCl is explained [3] as

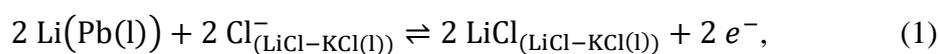

$$2\,\text{Li}(\text{Pb}(l)) + 2\,\text{Cl}^-_{(\text{LiCl}-\text{KCl}(l))} \rightleftharpoons 2\,\text{LiCl}_{(\text{LiCl}-\text{KCl}(l))} + 2\,e^-, \quad (1)$$

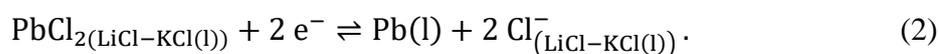

$$\text{PbCl}_{2(\text{LiCl}-\text{KCl}(l))} + 2\,e^- \rightleftharpoons \text{Pb}(l) + 2\,\text{Cl}^-_{(\text{LiCl}-\text{KCl}(l))}. \quad (2)$$

This leads to the overall cell displacement reaction:

$$2\,\text{Li}(\text{Pb}(l)) + \text{PbCl}_{2(\text{LiCl}-\text{KCl}(l))} \leftrightarrow \text{Pb}(l) + 2\,\text{LiCl}_{(\text{LiCl}-\text{KCl}(l))} \qquad (3)$$

The use of membranes is a widely accepted concept in the field of batteries and fuel cells. Mostly, ion-exchange membranes are used as an electrolyte barrier between the half-cells. The membranes fabricated for this study enable ion exchange because of their porous nature and are also electronically conductive to allow for a "faradaic protection" reaction. When $Pb^{2+}$ ions reach the lower surface of the membrane, they will be reduced faradaically into Pb metal. The lead droplets, formed at the surface of the membrane, sink back into the pool of Pb (l). This self-discharge reaction, which is caused by diffusion of lead ions from the bottom of the electrolyte towards the membrane, is of course detrimental. The membrane acts here simply as a lead barrier preventing irreversible lead transfer from the positive electrode to the negative electrode. LDBs built in the past exhibited a coulombic efficiency of 92% and an energy efficiency of 71% while operating at a current density of 150 mA cm$^{-2}$ at a temperature of 410 °C. The loss in coulombic efficiency is not caused by irreversible side reactions but rather simply by the self-discharge reaction, explained above [3]. LDBs are promising but require further mitigation strategies to reduce self-discharge. The latter is the scope of this study.

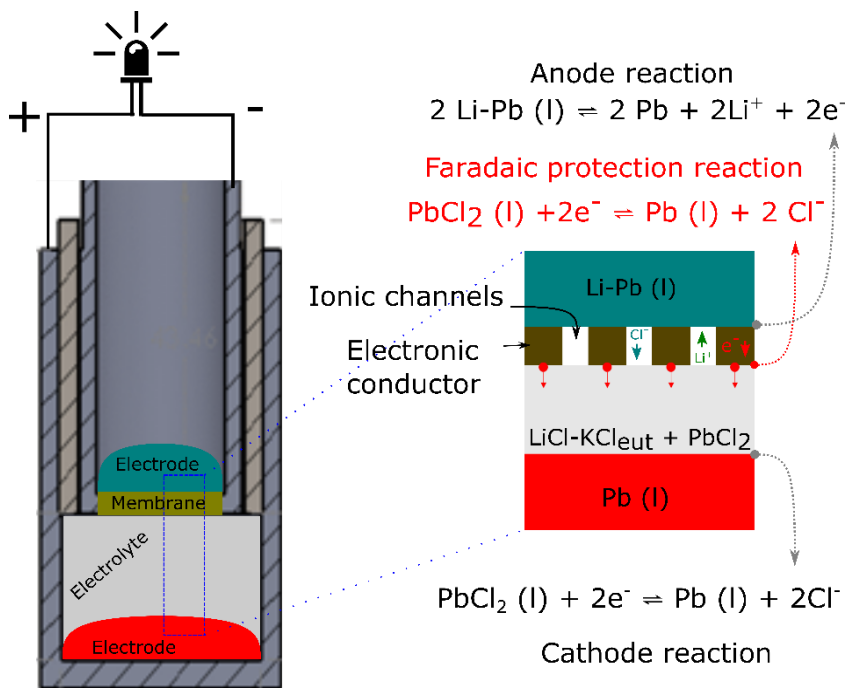

**Fig. 1.** Schematic of a liquid displacement battery. The detailed view shows the ion transport and faradaic reaction across a porous electronically conductive membrane.

While the reported battery chemistry is relatively recent, the electrochemical reactions explained above are well established and have been used for lead deposition from molten chlorides. The characteristics of the mixtures of molten salts and metals are reported elsewhere [5]. The diffusion coefficient of lead ions in molten chlorides has been reported to be $2 \times 10^{-5}$ cm$^2$ s$^{-1}$; the cathodic deposition of lead was found to be diffusion controlled [6–9] for the dilute solution of $PbCl_2$ in molten KCl-LiCl at 400 °C. A formulation based on applying Faraday's law and species transport equations led to a relationship highlighting the direct proportionality of the self-discharge current density to the diffusion coefficient. The mathematical formulation is given in the supplementary information.

Different remedies can be used to reduce self-discharge such as increasing the thickness of the electrolyte (salt) layer, reducing the solubility of $PbCl_2$ in the electrolyte, lower the operating temperature, freezing the electrolyte during cell idling, binding of $PbCl_2$ to some other chemical moiety inside the Pb, or by using an electrolyte featuring a miscibility gap. The idea of this study is to convert $PbCl_2$ into another complex compound leading to a reduction of the diffusion of the Pb-ions.

## 2. Materials and experimental methods

### 2.1. Chemicals

Lead (99.99%, anhydrous, Sigma Aldrich) was used in the electrodes. The electrolyte was prepared using the eutectic composition for a mixture of lithium chloride (LiCl, 99.9% anhydrous supplied by Alfa Aesar) and potassium chloride (KCl, 99.9% anhydrous supplied by Alfa Aesar). Membranes were prepared using titanium nitride (TiN) powder (Alfa Aesar, 99.7%, <10 μm) and magnesia (MgO) nano-powder (Inframat Advanced Materials). Lead-oxide was also obtained from Alfa Aesar.

### 2.2. Preparation of membrane

The porous TiN membrane referred to as the faradaic membrane was prepared using the method reported by [3]. The mixture having 5.4 g TiN powder and 0.6 g MgO nano-powder was mixed thoroughly by a pestle and then transferred into the graphite crucible (outer diameter: 2.54 cm, inner diameter: 2.3 cm, height: 4.4 cm) with an opening at the bottom having a diameter of 2 cm. This graphite crucible is acting as a current collector for the negative electrode as shown in Fig. 2. The powder inside the crucible was hydraulically pressed by inserting a graphite rod. This gives mechanical strength to the powder and helps in attaching it to the crucible surface and helps to avoid any leakage of the molten negative electrode during battery operation. The assembly was then transferred to the tube furnace which was purged three times with ultrapure Ar (99.99%) and then blanketed with argon gas and heated to 1100 °C at a ramping rate of 6 °C min$^{-1}$. The prepared assembly of the negative current collector carrying the mounted membrane was taken out of the furnace and allowed to cool to room temperature.

*2.3. Assembly and assessment of the liquid displacement cell*

The assembly of the liquid displacement cell is shown in Fig. 2. Graphite crucibles were used to contain the liquid metal electrode and molten salts. The positive electrode was connected to a metal rod serving as current collector. The crucibles were insulated by using a ceramic sleeve. In the beginning, the positive electrode was loaded with layers of lead granules followed by a mixture of LiCl & KCl as molten salt electrolyte having an eutectic composition [10]. The negative electrode consisting of lead granules was placed on top of the solid electrolyte. This assembly was placed inside the furnace maintained at 415 °C to obtain molten layers of all the added reagents. It was ensured that the bottom part of the negative electrode with its membrane was in good contact with the molten salt electrolyte before the start of measurements.

Data obtained from charge/discharge cycling enabled the calculation of coulombic, voltage, and energy efficiencies ($\eta_{CE}$, $\eta_{VE}$, and $\eta_{EE}$, respectively) for each cycle as

$$\eta_{CE_i} = \frac{\int_{t_{d,0}}^{t_{d,f}} I_d \mathrm{d}t}{\int_{t_{c,0}}^{t_{c,f}} I_c \mathrm{d}t} \quad , \tag{6}$$

$$\eta_{VE_i} = \frac{\int_{t_{d,0}}^{t_{d,f}} E_d(t) \mathrm{d}t}{\int_{t_{c,0}}^{t_{c,f}} E_c(t) \mathrm{d}t} \times \frac{(t_{c,f}-t_{c,0})}{(t_{d,f}-t_{d,0})} \quad , \tag{7}$$

$$\eta_{EE_i} = \eta_{CE_i} \times \eta_{VE_i} \quad , \tag{8}$$

where $I$ is the current, $E$ is the cell potential difference, $t$ is time, subscript $i$ stands for the cycle number, subscript c for the charge, subscript d for discharge, subscript 0 for initial, and subscript f for final.

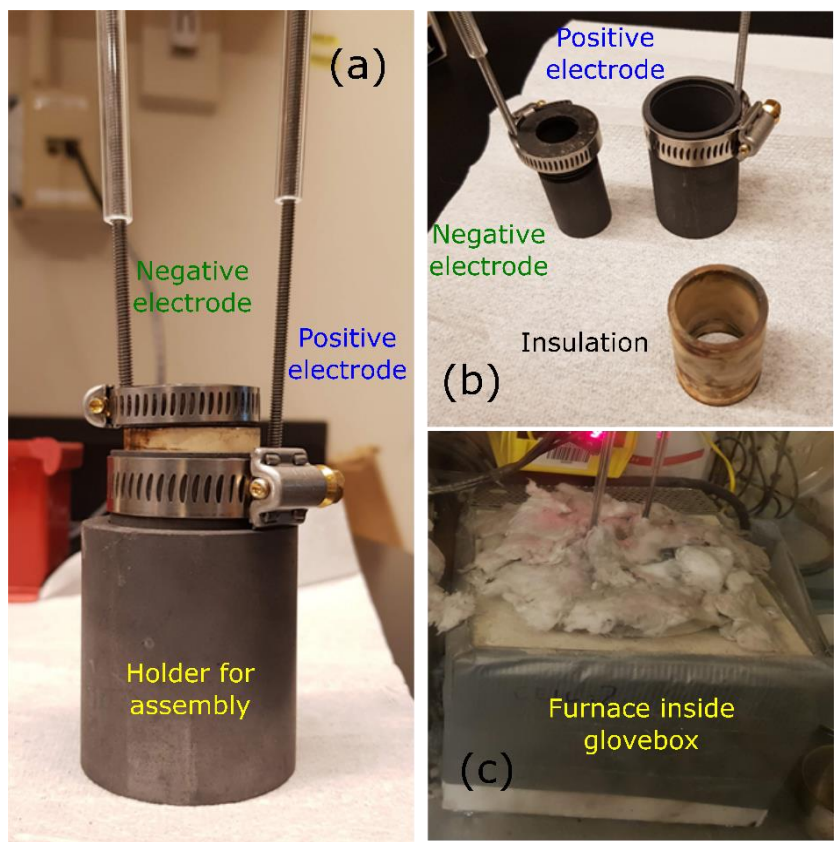

**Fig. 2.** Experimental setup of the liquid displacement battery fitted with the faradaic membrane. (a) Assembly of all the components. (b) Components. (c) Furnace inside the glovebox having the operated cell.

An activation of the cell was performed before conducting any measurements. Self-discharge was measured by charging the cell up to the capacity of 500 mAh and then recording the open circuit potential ($E_{OCP}$) at the no-load condition for up to 48 h. This technique helps to isolate the impact of self-discharge on the performance of the cell. Impedance measurements were recorded at 10 mV amplitude and across the frequency range of 2.5 kHz to 1 Hz using the Autolab workstation. Potentiostatic measurements were performed to acquire polarization data from 0.4 to 2.2 V at a scan rate of 10 mV s$^{-1}$. The ohmic resistance contribution from the wires – 0.02 Ω – was subtracted from all polarization and impedance measurements.

## 3. Results and discussion

### 3.1. Mitigation of self-discharge

The activation of the liquid metal electrodes and the molten salt electrolyte was performed through charge-discharge cycling (CDC) until the cell reached the operating capacity of 500 mAh. The cycling was done step-wise by operating the cell at the capacities of 25, 250, and 500 mAh as shown in Fig. 3 (i.e., CDC at 50, 100, and 200 mA for charging times of 0.5, 2.5, and 2.5 h, respectively). During the activation phase, the coulombic efficiency at 25 mAh capacity increased from 25.5% to 90.1% in 5 cycles, which indicates that there was no short-circuiting though the membrane. The responses of the cell recorded at 250 and 500 mAh have shown similar values of efficiency improvement. The maximum value of the coulombic efficiency achieved at 500 mAh was 91.3% and that of the voltage efficiency 74.98%, which served together with the state-of-the-art value [3] as a comparative benchmark.

The present study is a continuation of a previous article [3] which can be consulted for the detailed chemical and electrochemical characterizations of the lower surface of the membrane including scanning electron microscopy (SEM) images and energy dispersive X-ray spectroscopy (EDS) mapping performed before and after 100 cycles. The permeation rate of $Pb^{2+}$ was measured by direct-current plasma chemical analysis. It was also evident that the performance of the cell is stable until 400 cycles. Nevertheless, it was observed that the battery exhibited a certain self-discharge at open circuit conditions. Therefore, retention of the stored energy becomes the primary concern, and repetition of other results was forestalled to remain focused on the mitigation of self-discharge.

There was an interesting study reported by Pershin et al. [6] for the electrodeposition of Pb. They reported the formation of a $Pb_2OCl_2$ compound when adding PbO to a $PbCl_2$ melt. Hence, the addition of PbO to LiCl-KCl-$PbCl_2$ melt is expected to result in the reaction:

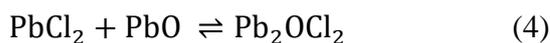

$$PbCl_2 + PbO \rightleftharpoons Pb_2OCl_2 \qquad (4)$$

The presence of PbO in the electrolyte complements the function of the faradaic membrane by providing additional steps for the electrochemical conversion of $Pb^{2+}$. In this way it serves as an additional barrier to the diffusion of $Pb^{2+}$. This leads to the modification in the positive electrode half-cell reaction and can be described as

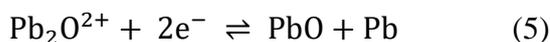

$$Pb_2O^{2+} + 2e^- \rightleftharpoons PbO + Pb \qquad (5)$$

The diffusion-controlled process of self-discharge can be decreased because of the reduced concentration of free $Pb^{2+}$ ions resulting from the formation of the $Pb_2O^{2+}$ ions in the electrolyte. These two additional side reactions complement the functioning of the faradaic membrane. Pershin et al. [6] also optimized the effect of PbO in the $PbCl_2$ melt by testing various molar percentages of 0.54%, 1.20%, 1.78%, and 2%. The best results were observed at 2% which was therefore chosen for further testing in the liquid metal displacement battery.

Self-discharge was examined by charging the cell up to the capacity of 500 mAh and then measuring the open circuit potential ($E_{OCP}$) for up to 48 h without any applied current. During this resting phase, the cell was maintained at constant temperature and other ambient conditions; therefore, any change in the value of $E_{OCP}$ could be attributed to the self-discharge of the cell.

The result of the self-discharge test as a function of $E_{OCP}$ is shown in Fig. 4. It was observed that the cell potential kept on decreasing over time and there is a rapid drop for the control experiment after 24 h of resting phase. However, a significant improvement was observed after the addition of 2 mol% PbO to the electrolyte. Then, the cell voltage remains stable even after 24 h and only a slight decrease was observed up to 48 h of resting phase. The addition of PbO to the electrolyte results in formation of a complex compound $Pb_2OCl_2$ and works as an additional filter to stop the crossover of $Pb^{2+}$ ions which are responsible for the self-discharge. Interestingly, the two curves overlay each other for the region of up to 24 h, and the rate of self-discharge did not get prominent until $E_{OCP} = 1.5$ V.

During discharging, the reduction of $Pb^{2+}$ is supported by oxidation of Li from the Li-Pb pool at the negative electrode. The pores of the membrane act as ionic channels and assist the passing of $Li^+$ into the electrolyte. Moreover, there are two pathways for the electrons and the value of the coulombic efficiency ($\eta_{CE} = 91\%$) as shown in Fig. 5 suggests that the dominant pathway is the external circuit taking most of the total output of current. The electronically conductive membrane is providing the second pathway which facilitates the reaction of $Pb^{2+}$ at the lower surface of the membrane thereby blocking the transport of $Pb^{2+}$ to the negative electrode (having a Li-Pb pool above). This drives the liquid Pb back to the positive electrode (present as a Pb pool on the bottom of the cell). Since the addition of PbO initiates additional reversible reactions, the need for electrons to be supplied through the second pathway is reduced. This in return increased the current flowing from the dominant pathway, i.e., external circuit, resulting in an increase in coulombic efficiency up to 97% without affecting the voltage efficiency as shown in Fig. 5(c).

Efficacy of this remedy was also tested for longer battery cycling and the cell was operated for 146 hours (*ca.* 6 days) to perform 30 charge-discharge cycles. The coulombic efficiency ($\eta_{CE}$) and discharge capacity of the cell was dropped to only 0.84 % and 1.82 % between the 5th and 30th cycle which indicate that the self-discharge of the cell is inhibited by the addition of PbO. The charge-discharge cycles, efficiencies, and discharge capacity for each cycle is shown in Fig. S 3 (a) & (b).

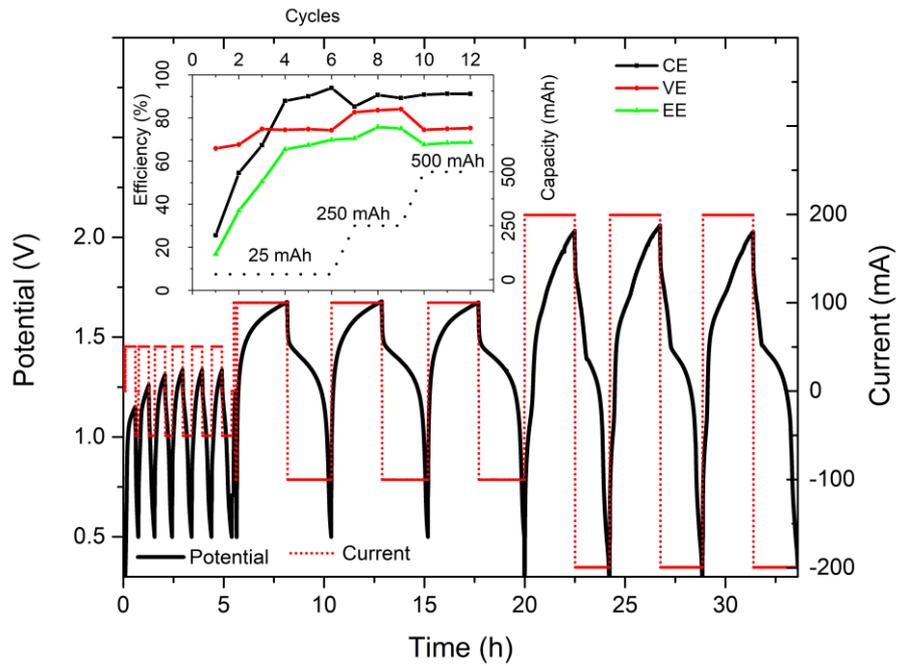

**Fig. 3.** Charge-discharge cycling for activation phase in three steps. Step 1: Capacity of 25 mAh at a constant current of 50 mA for 0.5 h. Step 2: Capacity of 250 mAh at a constant current of 100 mA for 2.5 h. Step 3: Capacity of 500 mAh at a constant current of 200 mA for 2.5 h (Inset shows $\eta_{CE}$, $\eta_{VE}$, $\eta_{EE}$, and capacity for the cycles respectively).

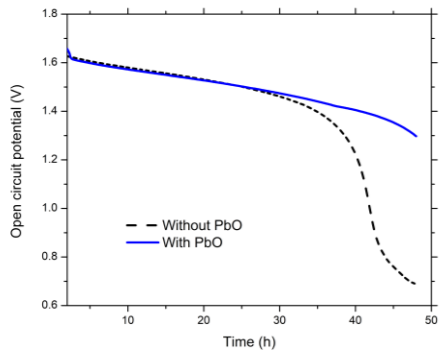

**Fig. 4.** The result of the self-discharge test without the addition of PbO and after the addition of 2 mol% PbO to the electrolyte.

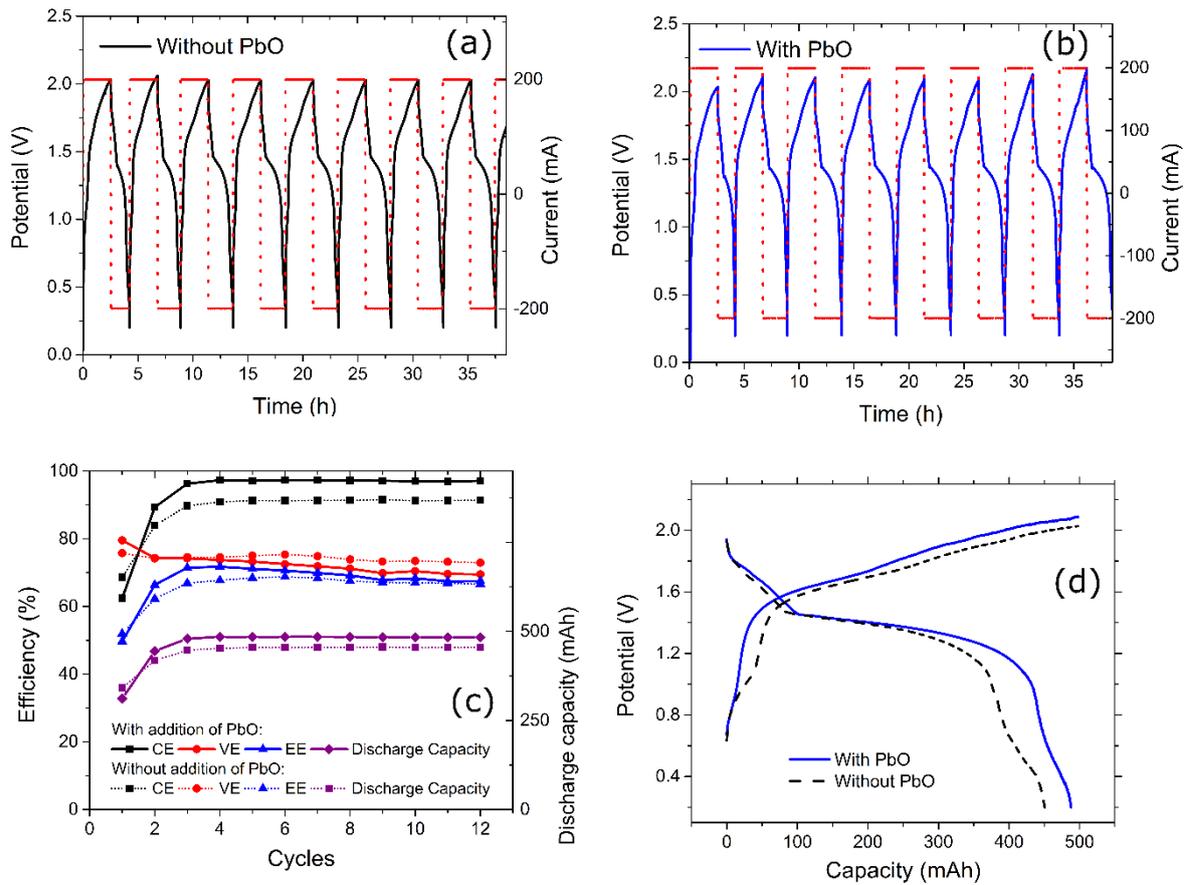

**Fig. 5.** Charge discharge cycling of the liquid displacement cell having a capacity of 500 mAh at a current of 200 mA. (a) Control experiment without the addition of PbO. (b) After addition of 2 mol% PbO to the electrolyte. (c) Efficiencies of both experiments. (d) Comparison of capacity vs. potential for the best performing charge-discharge cycle.

*3.2. Impact of mitigation on the impedance of the cell*

Electrochemical impedance spectroscopy (EIS) was used to further characterize the performance of the cell. A higher impedance of the cell results in lower voltage efficiency and, in turn, lower round-trip energy efficiency. It is important to quantify the overall impedance of the cell and then to further analyze it to get the contribution of each component. Initially, the impedance curve obtained from EIS measurements was fitted to a Randles circuit which helps to analyze the first semi-circle highlighting the ohmic resistance and charge-transfer resistance of the cell. The Randles circuit fittings were used as the starting point to get values for the model of the equivalent circuit. The impedance curve was further fitted to the equivalent circuit model considering the physical phenomena occurring in the cell.

Since existing reported results were not available, the EIS measurements were validated using the Kronig-Kramers test (K-K test) as shown in Fig. S 2. This test assesses the steady-state condition of the electrochemical system. If the investigated system is dynamic, i.e., changing with time, then this test fails. This can happen due to the effect of aging, temperature change, non-equilibrium initial state, etc. Failure of the K-K test usually means that no good fit can be obtained using the equivalent circuit method. The Kronig-Kramers test ensures causality, linearity, and stability: Here, causality signifies that the recorded response is related only to the excitation signal generated by the potentiostat during measurement. Linearity signifies that the recorded response is linear and the perturbation small enough to avoid any change in the open circuit potential of the cell during measurement. Finally, stability highlights that the electrochemical system is not changing the state-of-charge (SOC) with respect to time [11,12].

The measurements were successfully validated by the K-K test and fitted to the equivalent circuit as shown in Fig. 6. The inductance of wires between the cell, glove box, and potentiostat is denoted by $L_1$. The *x*-intercept of the Nyquist plot indicates the ohmic contribution of the cell and is denoted in the equivalent circuit by the resistive component $R_S$. The latter combines the ohmic resistances of the cathode, electrolyte, membrane, anode, and current collectors. The double-layer effect at the interfaces of the electrolyte with anode/cathode is modeled using an analogy of $RC_{CPE}$ circuits, where $C_{CPE}$ is a constant phase element (CPE). The fitted parameters can help to estimate the overall impedance of the cell by using the following equations, which are commonly applied for impedance analysis of the RC circuit (it is assumed that the constant phase element is behaving as a capacitor):

$$|Z_{RC,A}| = \frac{1}{\sqrt{\left(\frac{1}{R_A}\right)^2 + (\omega_A C_A)^2}}, \quad (9)$$

$$|Z_{RC,C}| = \frac{1}{\sqrt{\left(\frac{1}{R_C}\right)^2 + (\omega_C C_C)^2}}, \quad (10)$$

$$|Z_{CELL}| = R_S + |Z_{RC,A}| + |Z_{RC,C}|. \quad (11)$$

Here, subscript A refers to the anode and C to the cathode, $Z_{RC}$ is the impedance of the RC circuit, $Z_{CELL}$ is the estimated total impedance of the cell and $\omega$ is the frequency.

The equivalent circuit fits the individual impedance curves (Fig. 6) very well. The EIS parameters used in the equivalent circuit model were determined with good accuracy (Table 1). The errors for each parameter are within the 95% confidence interval according to the equivalent circuit analysis in ZView and lie within the Chi-squared criterion ($< 1 \times 10^{-3}$). There is a negligible effect of PbO addition on the ohmic resistance of the cell which indicates that neither the electrolyte layer thickness changes nor does the conductivity of the electrolyte. In comparison to this, the anodic impedance increased by ca. 32% (from 592 to 783 m$\Omega$) because the charge transfer resistance increased at the anode due to the additional side reactions. The cathodic impedance decreased, which is possibly associated with the physicochemical properties of the melt (LiCl-KCl-PbCl$_2$-Pb$_2$OCl$_2$) and also because of the availability of more Pb$^{2+}$ ions. Interestingly, this trade-off maintained the overall impedance of the cell, which is almost the same when adding PbO to the electrolyte. This finding is also supported by the earlier results showing that the coulombic efficiency increased without any effect on the voltage efficiency.

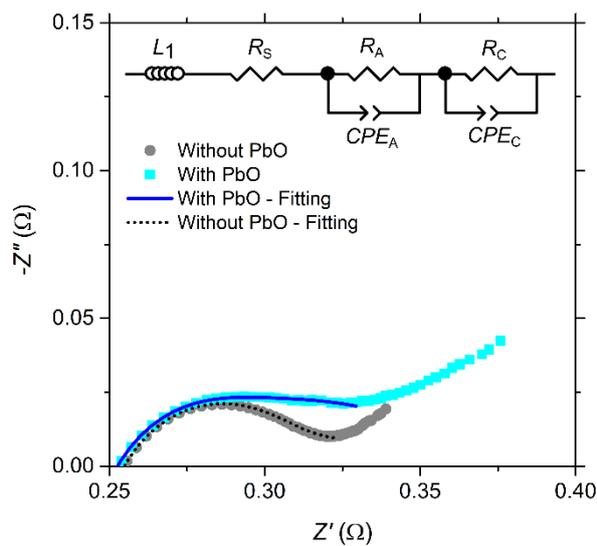

**Fig. 6.** Impedance curve obtained by electrochemical impedance spectroscopy in the frequency range from 2.5 kHz to 1 Hz having an amplitude of 10 mV. The inset shows the equivalent circuit model used for curve fitting.

**Table 1.** Fitting parameters and RC circuit impedance analysis of the equivalent circuit for the impedance curves shown in Fig. 6.

|  | Without PbO | | With PbO | |
|---|---|---|---|---|
| Elements | Value | Error (%) | Value | Error (%) |
|  | Fitted parameters | | | |
| $L_1$ (mH) | $2.16 \times 10^{-3}$ | 0.37 | $2.08 \times 10^{-3}$ | 1.12 |
| $R_S$ (Ω) | $2.56 \times 10^{-1}$ | 0.02 | $2.54 \times 10^{-1}$ | 0.05 |
| $R_A$ (Ω) | $6.48 \times 10^{-2}$ | 0.55 | $8.46 \times 10^{-2}$ | 0.84 |
| $CPE_A$ (mF) | $4.65 \times 10^2$ | 0.29 | $4.69 \times 10^1$ | 0.65 |
| $R_C$ (Ω) | $6.59 \times 10^{-2}$ | 0.08 | $5.31 \times 10^{-2}$ | 0.31 |
| $CPE_C$ (mF) | 2.99 | 0.21 | 2.77 | 0.8 |
| Chi-squared (Criteria < $1 \times 10^{-3}$) | $5.7 \times 10^{-7}$ | | $5.3 \times 10^{-6}$ | |
| | RC-circuit impedance analysis | | | |
| $|Z_{RC,A}|$ (Ω) | $5.92 \times 10^{-4}$ | | $7.83 \times 10^{-3}$ | |
| $|Z_{RC,C}|$ (Ω) | $6.58 \times 10^{-2}$ | | $5.31 \times 10^{-2}$ | |
| $|Z_{CELL}|$ (Ω) | $3.22 \times 10^{-1}$ | | $3.15 \times 10^{-1}$ | |

*3.3. Impact of mitigation on the polarization of the cell*

The polarization curve analysis is based on the standard electrochemical technique of linear sweep voltammetry (LSV) for characterizing the performance of the battery and provides information about the performance losses in the cell under operating conditions. The generalized polarization curve of a battery is given in supplementary information as Fig. S 1. The peaks of the polarization curve highlight the limiting current density and limiting power that can be achieved from that particular cell. The curve obtained while discharging the cell gives us the useful parameters of activation overpotentials ($\eta_a$), ohmic losses ($\eta_o$) and mass transport ($\eta_m$) overpotentials. These values can be extracted using the information of the ohmic resistance from the impedance analysis.

The polarization curve obtained while charging the cell from 0.4 to 2.2 V at a scan rate of 10 mV s$^{-1}$ is shown in Fig. 7. The control experiment (without PbO) achieved a limiting current density of 575 mA cm$^{-2}$ and limiting power up to 2.53 W whereas the remedy experiment (with PbO) achieved a limiting current density of 831 mA cm$^{-2}$ and a limiting power of 3.66 W. Moreover, the polarization curve was obtained while discharging the cell from the SOC = 80% to SOC = 20%, which is the range of varying the potentials from 1.8 to 0.3 V. The activation overpotential is negligible because of the fast kinetics at liquid-liquid interfaces. Ohmic resistance values obtained from EIS were used to get iR-free curves which help to quantify the ohmic overpotential ($\eta_o$); the remaining curve is the effect of the mass-transport overpotential ($\eta_m$). The values of the various overpotentials are shown in Fig. 8. The gains in the performance of the cell are also evident through these results.

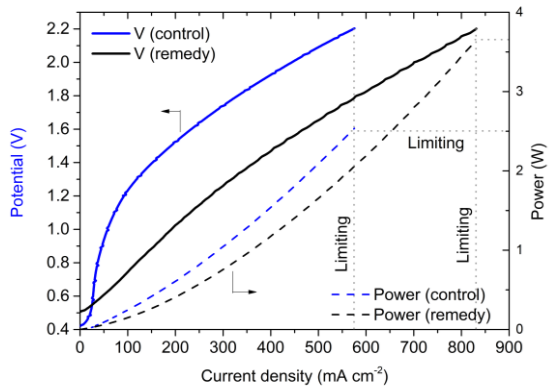

**Fig. 7.** Polarization curves while charging the cell from 0.4 to 2.2 V. The solid lines represent voltage as a function of current density and the dotted lines represent power as a function of current density (mA cm$^{-2}$). The active area of the cell is 2 cm$^2$.

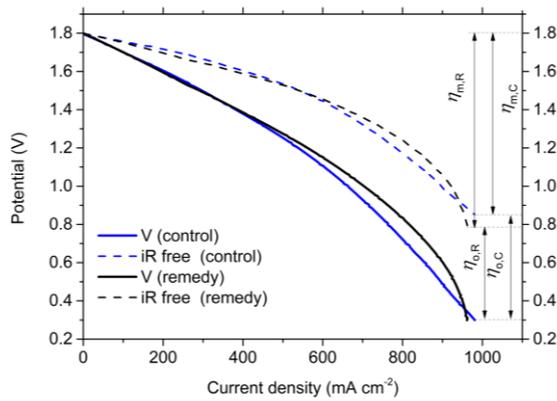

**Fig. 8.** Polarization curves while discharging the cell from 1.8 to 0.3 V at a scan rate of 10 mV s$^{-1}$. The solid lines represent voltage as a function of current density and the dotted lines represent iR-free curves.

## 4. Conclusions

A formulation involving Faraday's law and the Nernst-Planck equation leads to a relationship that indicates that the self-discharge current density is directly proportional to the diffusion coefficient. This directs this study to the solution of adding lead oxide (PbO) to the electrolyte so as to reduce the diffusion coefficient as well as the Pb ion concentration by initiating two additional side reactions. The limiting current density and power were improved from 575 to 831 mA cm$^{-2}$ and from 2.53 to 3.66 W, respectively. The addition of PbO decreases the ohmic losses, but the mass transport overpotentials increased due to the

14/20

effect of PbO on the diffusion of $Pb^{2+}$ in the electrolyte. The equivalent circuit model estimated the contribution of impedance by contacts and interfaces. The remedial actions reported here resulted in the highest reported coulombic efficiency for liquid displacement batteries, i.e., an increase from 91% to 97% without affecting the voltage efficiency. The addition of PbO slows the diffusion-controlled self-discharge process by complementing the faradaic function of the membrane with additional reversible side reactions.

**Acknowledgments**

This work was financially supported by the research unit of Group Sadoway Laboratory, Department of Materials Science and Engineering, Massachusetts Institute of Technology, 77 Massachusetts Avenue, Cambridge, MA, 02139-4307, United States. Kashif Mushtaq is grateful to the Portuguese Foundation for Science and Technology (FCT) for his Ph.D. scholarship (PD/BD/128041/2016). This work was also financially supported by: Base Funding - UIDB/00511/2020 of the Laboratory for Process Engineering, Environment, Biotechnology and Energy – LEPABE - funded by national funds through the FCT/MCTES (PIDDAC). The authors acknowledge continuous support from MIT Portugal Program.

**References**

[1]     K. Mushtaq, T. Lagarteira, A.A. Zaidi, A. Mendes, J. Energy Storage 40 (2021) 102713–102722.

[2]     W. Zhang, D. Wang, W. Zheng, J. Energy Chem. 41 (2020) 100–106.

[3]     H. Yin, B. Chung, F. Chen, T. Ouchi, J. Zhao, N. Tanaka, D.R. Sadoway, Nat. Energy 3 (2018) 127–131.

[4]     J. Figgener, P. Stenzel, K.P. Kairies, J. Linßen, D. Haberschusz, O. Wessels, G. Angenendt, M. Robinius, D. Stolten, D.U. Sauer, J. Energy Storage 29 (2020) 101153–101172.

[5]     V.M.B. Nunes, C.S. Queirós, M.J.V. Lourenço, F.J.V. Santos, C.A. Nieto de Castro, Appl. Energy 183 (2016) 603–611.

[6]     P. Pershin, Y. Khalimullina, P. Arkhipov, Y. Zaikov, J. Electrochem. Soc. 161 (2014) D824–D830.

[7]     A. Kisza, J. Kazmierczak, B. Børresen, G.M. Haarberg, R. Tunold, J. Electrochem. Soc. 144 (1997) 1646–1651.

[8]     B. Bo⁄rresen, G.M. Haarberg, R. Tunold, O. Wallevik, J. Electrochem. Soc. 140 (1993) L99–L100.

[9]  A. Kisza, J. Kaźmierczak, B. Børresen, G.M. Haarberg, R. Tunold, J. Appl. Electrochem. 25 (1995) 940–946.

[10] M. Hino, M.G. Kelly, J.M. Toguri, Can. Metall. Q. 29 (1990) 177–184.

[11] B.A. Boukamp, J. Electrochem. Soc. 142 (1995) 1885–1894.

[12] M. Urquidi-Macdonald, S. Real, D.D. Macdonald, Electrochim. Acta 35 (1990) 1559–1566.

[13] A.S. Basin, A.B. Kaplun, A.B. Meshalkin, N.F. Uvarov, Russ. J. Inorg. Chem. 53 (2008) 1509–1511.

[14] D. Aaron, Z. Tang, A.B. Papandrew, T.A. Zawodzinski, J. Appl. Electrochem. 41 (2011) 1175–1182.

**Graphical abstract**

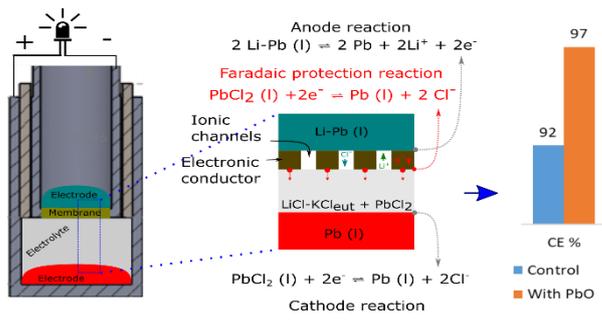

The addition of PbO acts as a mitigator to suppress self-discharge and enhance the performance of the liquid metal displacement battery.

**Supplementary Information**

**Self-discharge current density:**

A formulation based on Faraday's law and by applying the Nernst-Planck equation leads to a relationship highlighting the direct proportionality of self-discharge current and diffusion. The detailed derivation is given below. Faraday's law relates the amount of an active substance to the current as

$$n = \frac{It}{zF}, \quad (1)$$

where $n$ is the amount of substance, $I$ is an electrical current, $t$ is the total time of the applied constant current, $z$ is the electrons transferred per ion and $F$ is the Faraday constant. Considering the rate of reaction

$$\dot{n} = \frac{I}{zF} \quad (2)$$

and normalizing it by the surface area ($A$), becomes

$$N_i = \frac{\partial \dot{n}}{\partial A} = \frac{J}{zF}, \quad (3)$$

where $J$ is the current density, subscript $i$ refers to the active species (Pb in this case) and $N_i$ denotes the net flux of $Pb^{2+}$ without $Cl^-$. This leads to

$$J = F \sum_i z_i N_i. \quad (4)$$

According to the assumptions of negligible migration or convection, the transport equation for the active species reads

$$N = -D \nabla c_{Pb}. \quad (5)$$

Thus, the self-discharge current density is simply defined as

$$J = -zFD \nabla c_{Pb}, \quad (6)$$

where $c_{Pb}$ denotes the molar concentration of $Pb^{2+}$.

**Battery testing:**

A graphite crucible containing 7.7 g of Pb was used as positive electrode, and 26.9 g lead contained in the TiN membrane fitted graphite crucible were used as negative electrode. A MgO cylinder was used as insulation between both crucibles. The electrolyte was prepared using anhydrous lithium chloride (LiCl) and potassium chloride (KCl). The eutectic composition (for Li/K = 0.592:0.408 [13]) allowed to decrease the melting point of the electrolyte mixture and the temperature of the cell was maintained at 410 °C measured by an ASTM type-K thermocouple. The distance between the TiN membrane and bottom of the positive graphite crucible was 2 cm.

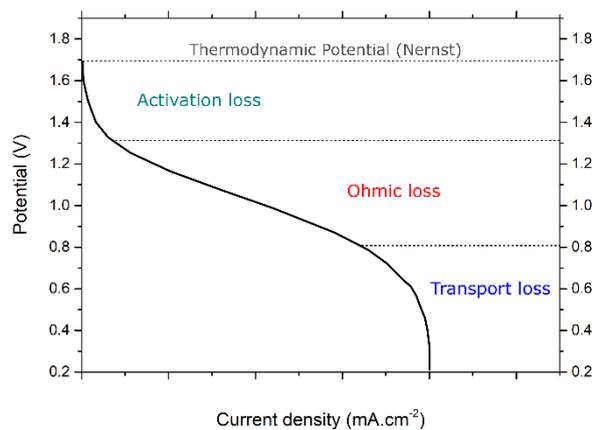

Fig. S 1. Generalized polarization curve for a battery indicating the dominant source of the overpotentials in each region [14].

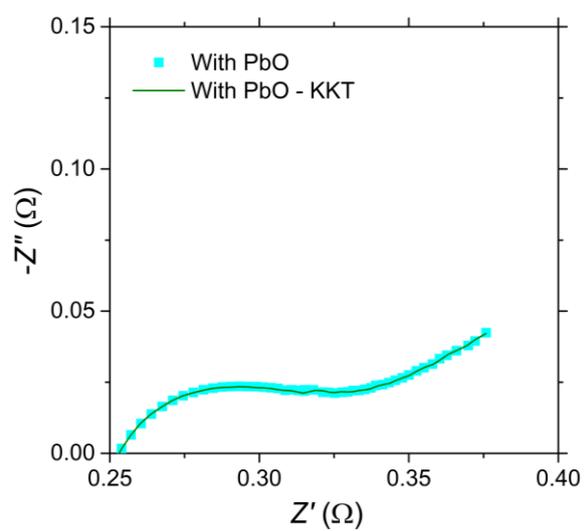

Fig. S 2. Kramers-Kronig transforms for the cell having addition of PbO.

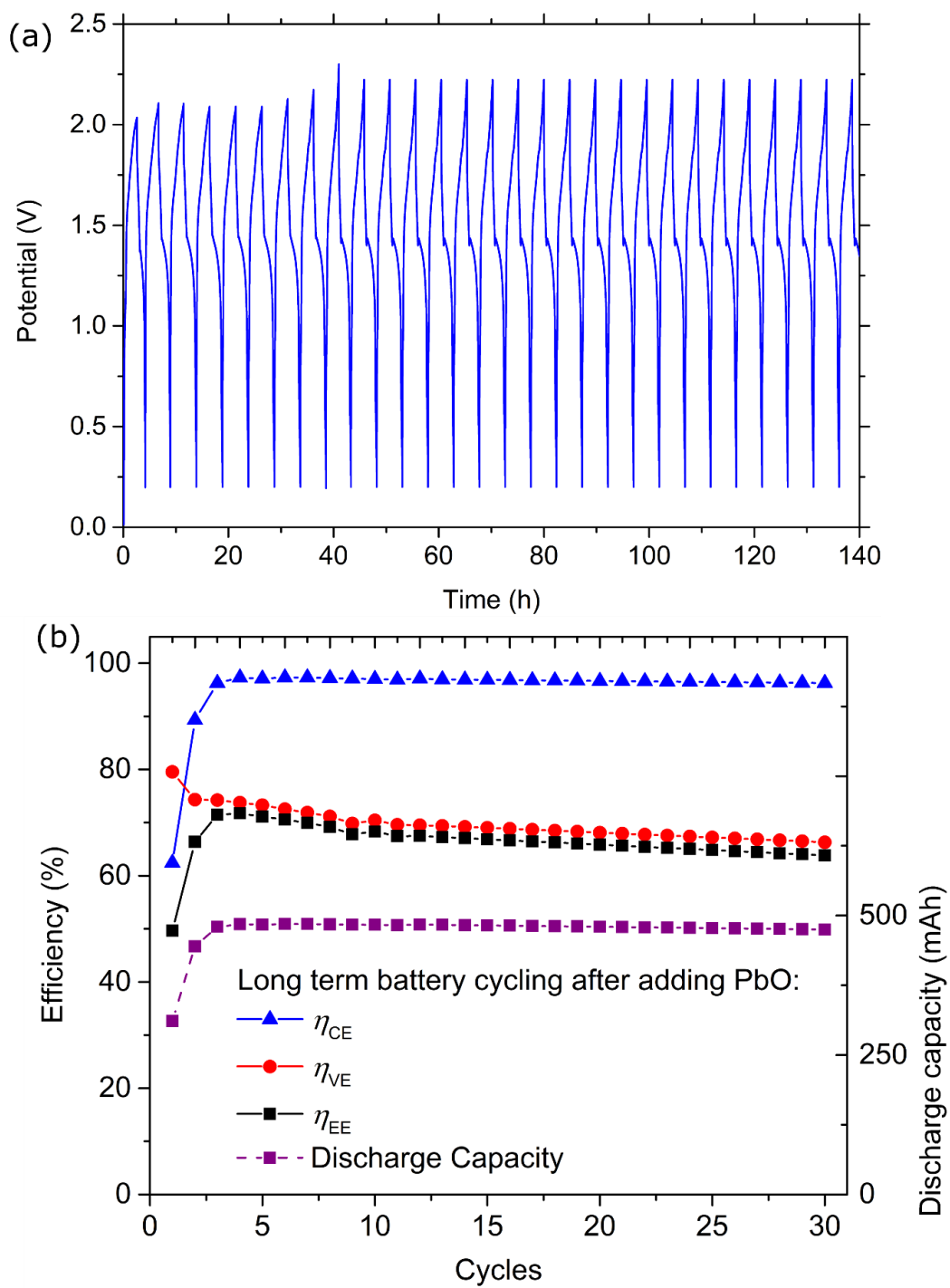

Fig. S 3. Liquid metal displacement cell having remedy of PbO addition to indicate the stability of the cell cycling for longer time. (a) It was operated for 146 hours (*ca.* 6 days) to perform 30 charge-discharge cycles. (b) Efficiency and discharge capacity